\documentclass[twocolumn,aps,pre,longbibliography]{revtex4-2}
\usepackage{epsfig}
\usepackage{natbib}
\usepackage{ulem}
\usepackage{xspace}
\usepackage{xcolor}
\usepackage{bbold}
\usepackage{caption}
\usepackage[title,toc,titletoc]{appendix}
\usepackage{float}
\usepackage{subcaption}
\usepackage[left=2.00cm, right=2.00cm, top=3.00cm,bottom=3.00cm]{geometry}
\usepackage{amsmath}
\usepackage{amsfonts}
\usepackage{amssymb}
\usepackage{graphicx}
\usepackage{listings}
\usepackage{mathtools}
\usepackage{amsbsy}
\usepackage{ifpdf} 
\usepackage{dsfont}
\usepackage{multirow}
\usepackage{xparse}
\usepackage{array}
\usepackage{etoolbox}
\usepackage{setspace}
\usepackage{physics}
\usepackage{titlesec} 
\usepackage{fancyhdr} 
\usepackage[colorlinks=true,
urlcolor=blue,
linkcolor=blue,
anchorcolor=blue,
citecolor=blue
]{hyperref}

\begin{document}  
	 \author{Avraham Samama$^1$}
	\author{Eli Barkai$^1$}
	\affiliation{$^1$Department of Physics, Institute of Nanotechnology and Advanced Materials, Bar-Ilan University, Ramat-Gan, 52900, Israel}
	\title{Statistics of Long-Range Force Fields in Random Environments: Beyond Holtsmark}
	\begin{abstract}
Since the times of Holtsmark (1911), statistics of fields in random environments have been widely studied, for example in astrophysics, active matter, and line-shape broadening. The power-law decay of the two-body interaction, of the form $1/|r|^\delta$, and assuming spatial uniformity of the medium particles exerting the forces, imply that the fields are fat-tailed distributed, and in general are described by stable Lévy distributions. With this widely used framework, the variance of the field diverges, which is non-physical, due to finite size cutoffs. We find a complementary statistical law to the Lévy-Holtsmark distribution describing the large fields in the problem, which is related to the finite size of the tracer particle. We discover bi-scaling, with a sharp statistical transition of the force moments taking place when the order of the moment is $d/\delta$, where $d$ is the dimension. The high-order moments, including the variance, are described by the framework presented in this paper, which is expected to hold for many systems. The new scaling solution found here is non-normalized similar to infinite invariant densities found in dynamical systems.
	\end{abstract}

	\maketitle	
\section{INTRODUCTION}
In 1919, Holtsmark considered the problem of the distribution of force fields in the context of the chaotic motion of charged particles in a plasma \cite{Jean}. Similarly, Chandrasekhar and Von Neumann examined the distribution of gravitational forces in the universe \cite{Von}. The basic question studied was the following: in an infinite system/universe with uniformly distributed charges/masses, what is the distribution of forces projected along the $z$-axis, $F_z$, acting on a tracer located on the origin \cite{Chandrash}? This distribution peaks at $F_z=0$, and its mean is zero due to symmetry, however the interesting aspect of the solution is that the variance of the field diverges, which is argued to be unphysical (see below) \cite{Jean,Von,Chandrash,Piet,Figueiredo,Wes,Dunkel,Gel}. Mathematically, Holtsmark's problem is related to the generalized central limit theorem \cite{Klafter}, \textit{i.e.} L\'{e}vy statistics, in which the distribution is fat-tailed \cite{Kolmogorov,Hazut}. The full connection to L\'{e}vy's stable laws, with truly vast applications \cite{Zarfaty,Sire,Bruce,Biel,Kuri}, is only seen by extending the original works \cite{Jean,Von} to include other force fields beyond Coulomb and Newton's gravitation law (see below). The statistical law discovered by Holtsmark and others is related to the power-law decay in space of forces acting between two bodies \cite{Chandrash}. Hence, the applications of this basic model and its extensions are found in many fields, for example, plasma physics \cite{Plasma}, astrophysics \cite{Chandrash,Von,finiteuniverse}, swimming micro-organisms \cite{Dais}, glassy systems \cite{Kador,Barkai}, forces in systems composed of dipoles \cite{Wes,Gel}, NMR \cite{Stoneham}, Olbers paradox \cite{finiteuniverse}, and in-homogeneous line-shape broadening \cite{Davidson}.

Notwithstanding previous works, here we introduce a complementary statistical law to those famous problems. Strictly speaking, the diverging variance of the force field is unphysical, as the original theory neglects an important excluded volume effect, namely originally the size of the tracer is taken to be zero. Mathematically, one goal of this letter is to use tools from infinite ergodic theory \cite{infdens,Erez,Davidson,Wan,Kesler,EliB,Leibovich,Ralf,Radons} to find a complementary statistical law for the force distribution. At the center of infinite ergodic theory stands the infinite invariant density \cite{Aaronson,Invariant}, which is a non-normalized function, hence its name. We show how this function can be used to describe the statistics of the force field, when the tracer size is finite and the density of bath particles is low. Importantly, in the field of active matter, this tool describes the largest forces in the problem, and these are crucial for current-day studies.

Recently, considerable work has been devoted to active transport, for example, self-propelled colloids, and biological swimming microorganism \cite{Micro,Kana}. The phenomenology of these systems is extremely rich, but one aspect of the problem is the ``static force fields akin to Holtsmark distribution" \cite{Jansen,Kuri,Dunkel,Dais} that in turn controls the dynamical features of the motion. The interaction is mediated by long-ranged hydro-dynamical dipole force fields \cite{Kana}, but it has a cutoff length scale $a$ defined below, just like other realistic forces. This cutoff scale is of great importance, as the forces the tracer experiences cannot be arbitrarily large. Hence, as mentioned, the Hotlsmark approach that yields a diverging variance of the force needs modifications. While the standard treatment assumes mono-scaling, namely, that the statistics of the force field is determined by a single scale, which is the Holtsmark scale defined below, we will show how the field distribution exhibits bi-scaling \cite{Strong-anomalous,anomalous in living cell,Displacement}, accompanied by a sharp statistical transition. This study is important for a vast number of systems \cite{Chandrash, Piet,Wes,Von,Figueiredo,Dunkel}, and with some modifications for ones driven by long-range active forces \cite{Martin,Kana}. Simply enough, the infinite density found in this letter describes the statistical properties of the largest forces in the problem, and these are important in the study of extreme events in many systems.

\section{MODEL}
Consider a rigid tracer with a radius $a$ that is centered in a sphere with a volume $V\rightarrow\infty$ in dimension $d=2$, or $d=3$. There are $N\rightarrow\infty$ randomly uniformly distributed bath particles inside the system, resulting in a finite overall density, $\rho=N/V$~\cite{Sire,Piet,Figueiredo,Chandrash}. The bath particles are treated as size-less charges, dipoles, masses, etc., and they cannot overlap with the rigid body on the origin. This in turn implies that we are considering the low density limit of the model where spatial correlations in the bath are neglected. Each particle applies a force on the tracer that decays like a power-law with the distance. The Cartesian axes in $d=3$ are denoted as $(x,y,z)$ and in $d=2$, $(x,z)$, so the total force toward the $z$-axis, denoted as $F_z$, is obtained by adding all the $z$ components of the forces applied on the tracer by all the $N$ particles,
\begin{equation}
	F_z=\sum_{i=1}^{N}(F_i)_z=\sum_{i=1}^{N}\frac{C\cos(\theta_i)}{|\mathbf{r}_i|^\delta},
	\label{3}
\end{equation}
where $C$ is a constant determined by the type of the particles. For Coulomb force, $\delta=2$ and $C=q^2/4\pi\epsilon_0\epsilon_r$, where $\epsilon_r$ is the dielectric constant in a medium, and $\epsilon_0$ is the one in vacuum. The opening angles from the $z$-axis and the distance from the tracer to particle $i$ are denoted by $\theta_i$ and $r_i$, respectively, and the force-law exponent is denoted by $\delta$.

The force $F_z$ is clearly a random variable, as it is a sum of many random contributions. The basic question is what is the force probability density function (PDF), $P(F_z,\xi)$, where we define $\xi = (\rho^{1/d} a)^{\delta}$? In particular we focus on the limit of small $\xi$, and the new finding of our work deal mainly with the large forces, as mentioned in the introduction. An explanation about the simulation of the model is provided in Appendix \ref{Appendix A}.

\section{CHARACTERISTIC FUNCTION AND MOMENTS}

This problem contains two force scales: $F_c=C/a^\delta $ which is the maximal force exerted on the tracer by a single bath particle, (the subscript ``$c$" stands for cut-off), as one can see from Eq.~\eqref{3} by simply inserting $r=a$ and $\theta=0$. The second is $F_H=\rho^{\delta/d} C $, which is the force scale studied by Holtsmark and others. The pair of force scales and connection
of the problem to fat tailed distributions, with cutoffs, imply that we can use bi-scaling ideas \cite{Strong-anomalous,RebePRL} namely, we will find two limiting laws for the distribution of forces. The relationship between $F_H$ and $F_c$, is characterized by
\begin{eqnarray}
	\xi^\alpha = \left(\frac{F_H}{F_c}\right)^\alpha=\rho a^d, {\rm where} \quad \alpha=\frac{d}{\delta}<2
	\label{17}
\end{eqnarray} 
\begin{figure}[H]
	\includegraphics[width=0.51\textwidth]{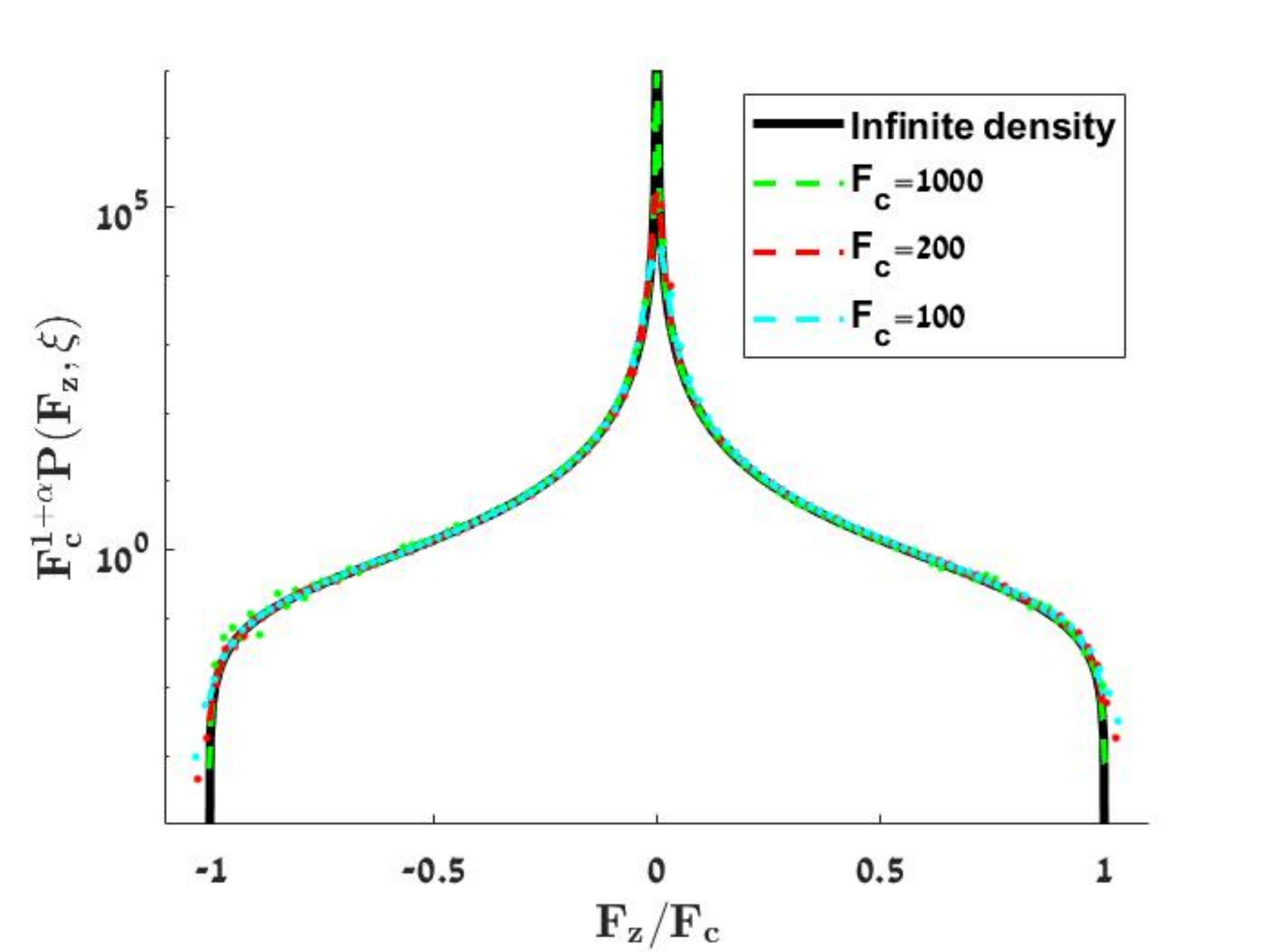}
	\caption{ $F_c^{1+\alpha}P(F_z,\xi)$ versus $\tilde{F}=F_z/F_c$ obtained from numerical simulation (dotted line) and compared with the infinite density (solid line) and the inverse Fourier transform of the characteristic function (dashed-line) found in Eq.~\eqref{44} and Eq.~\eqref{d3Characteristic} for $\alpha=3/2$, respectively. Clearly the infinite density is indeed a complementary statistical law, indicating that the Holtsmark's law is only part of the story. In particular the infinite density has a cut-off at $|\tilde{F}|=1$, which refers to the rare events. Thus, this law describes the large forces in the problem. For small $\tilde{F}$, the solution is non-integrable at the origin in the limit $F_c\rightarrow\infty$. We chose $C=1$, such that $F_c=1/a^\delta$, the density $\rho=3/4\pi$, the number of particles $N= 8000$, and the dimension $d=3$. }
	\label{fig:1}
\end{figure}
\noindent
and our interest as mentioned is in the limit of small $\xi$. Consider the characteristic function of $F_z$, given by
 \begin{equation}
 	\langle e^{ikF_z} \rangle =\left[\frac{\int_{V}dV \exp\left(ik\frac{C\cos(\theta)}{|\mathbf{r}|^{\delta}}\right)}{V}\right]^N,
 \end{equation}
 where we used Eq.~\eqref{3} and the fact that the bath particles are uniformly distributed in space. Since $V\rightarrow\infty$ and $N\rightarrow\infty$, we must interpret this integral. Therefore, we apply the following trick of adding and subtracting 1, so the characteristic function, also known as the Fourier transform of $P(F_z,\xi)$, denoted
 \begin{equation}
 \langle e^{ikF_z}\rangle=\int_{-\infty}^{\infty}P(F_z,\xi)e^{ikF_z}dF_z,
 \end{equation}
 is written as \cite{Stoneham}
\begin{eqnarray}
	\label{4}
	\langle e^{ikF_z}\rangle&=&\exp\left[-\rho\Omega_d\int_{0}^{\theta_d}f_d(\theta)d\theta \right. \\ \nonumber &~&  \left. \int_{a}^{\infty}\left(1-e^{ik\frac{C\cos(\theta)}{r^\delta}}\right)r^{d-1}dr\right],
\end{eqnarray}
where we used the mentioned excluded volume effect. $\Omega_d$ is the result of the integral over the azimuth angle in dimension $d$, such that $\Omega_2=1$, and $\Omega_3=2\pi$. $\theta_d$ is the polar angle, hence $\theta_2=2\pi$, and $\theta_3=\pi$. $f_2(\theta)=1$, and $f_3(\theta)=\sin(\theta)$.

\subsection{The infinite density for dimension three}
We first study the $d=3$ case. Using Eq.~\eqref{4}, the characteristic function for this case is
\begin{align}
	\label{d3Characteristic}
	&\langle e^{ikF_z}\rangle= \nonumber\\
	&\exp\left[-\frac{4\pi\xi^\alpha\left[{}_1F_2\left(-\frac{\alpha}{2};1-\frac{\alpha}{2},\frac{3}{2};-\frac{|F_ck|^2}{4}\right)-1\right]}{3}\right],
\end{align}
where ${}_1F_2$ is the generalized hypergeometric function~\cite{Hyper}. The moments of $F_z$ can be determined using the following equation:
\begin{equation}
\langle F_z^{2n}\rangle=(-1)^n\left.\frac{d^{2n}\langle e^{ikF_z}\rangle}{dk^{2n}}\right|_{k=0}
\label{MGF}
\end{equation}
After performing straightforward calculations (refer to Appendix \ref{Appendix B} for details), the variance can be expressed as:
\begin{equation}
	\langle F_z^2\rangle=\frac{4\pi \alpha \xi^\alpha F_c^{2}}{9(2-\alpha)},
	\label{F2d3}
\end{equation}
Thus, the cutoff force scale, $F_c$, governs the behavior of the variance. Similarly, the fourth moment is
\begin{eqnarray}
	\langle F_z^4 \rangle = F_c^4\left(\frac{4\pi \alpha \xi^\alpha}{15(4-\alpha)}+3\left(\frac{4\pi \alpha \xi^\alpha}{9(2-\alpha)}\right)^2\right),
	\label{F4d3}
\end{eqnarray}
and for $\xi\ll1$, the leading term of the fourth moment is of order of $\xi^\alpha$. This is not a coincidence, since the leading term of the sixth moment,
\begin{equation}
	\langle F_z^6 \rangle =F_c^6\left(\frac{4\pi \alpha \xi^\alpha}{21(6-\alpha)}+\frac{16\pi^2 \alpha^2 \xi^{2\alpha}}{9(2-\alpha)(4-\alpha)}+15\left(\frac{4\pi \alpha \xi^\alpha}{9(2-\alpha)}\right)^3\right),
	\label{F6d3}
\end{equation}
has an asymptotic behavior of $\xi^\alpha$ for $\xi\ll1$. The same can be done for the $2n$-th moment. Hence, for a non-negative integer $n$, the $2n-th$ moment for $\xi\ll1$ is
\begin{equation}
	\langle F_z^{2n}\rangle \sim \frac{4\pi\alpha \xi^\alpha}{3(2n-\alpha)(2n+1)} F_c^{2n}.
\label{moments}
\end{equation}
As discussed in the introduction, the second moment $\langle F_z^2\rangle$ diverges as $F_c\rightarrow\infty$, but $F_c$ is finite as long as $a > 0$, so moments of the force field do not diverge. Odd moments are zero due to symmetry. Notice that, the expression in Eq.~\eqref{moments} diverges if we analytically continue $n$ and set it to approach $\alpha/2$ ($n\rightarrow\alpha/2$) from above, and if we set $n=0$, we get a negative result, which violates the normalization condition. 

\section{INFINITE DENSITY}

Our next goal is to find a function that generates the moments presented in Eq.~\eqref{moments}. This function is called the infinite density. Using Eq.~\eqref{moments} and $[(2n-\alpha)(2n+1)]=[(2n-\alpha)^{-1}-(2n+1)^{-1}]/(1+\alpha)$, we search for a non-negative symmetric function $P_A(F_z,\xi)$ such that
\begin{eqnarray}
	\langle F_z^{2n}\rangle_A&=&2\int_{0}^{\infty}F_z^{2n}P_A(F_z,\xi)dF_z \nonumber\\&=&\frac{4\pi\alpha\xi^\alpha F_c^{2n}}{3(1+\alpha)}\left[\frac{1}{2n-\alpha}-\frac{1}{2n+1}\right],
	\label{SeperateMoment}
\end{eqnarray}
where $``A"$ stands for asymptotic, since the approach is valid for $\xi\ll1$. Naively, $P_A(F_z,\xi)$ is a PDF since it gives the moments of the force field, but this as we show soon is simply wrong.  Recall the Mellin transform \cite{MellinTrans} for a polynomial function, $f(x)$, that satisfies $f(|x|<1) = \left| x \right|^b$ and otherwise zero,
\begin{equation}
	\{Mf\}(s)\equiv\int_{0}^{\infty}x^{s-1}f(x)dx=\frac{1}{s+b}.
	\label{Mellin}
\end{equation}
Using Eq.~\eqref{SeperateMoment}, we see that in our case $s=2n+1$. Therefore, by applying the inverse Mellin transform on Eq.~\eqref{SeperateMoment}, we find that
\begin{align}
	&P_A(F_z,\xi)=\nonumber\\&
	\begin{cases}
		\frac{2\pi \alpha F_H^\alpha}{3(1+\alpha)F_c^{1+\alpha}}\left[\left(\frac{|F_z|}{F_c}\right)^{-1-\alpha}-1\right], & |F_z|<F_c\\
		0, & |F_z|\geq F_c.
	\end{cases}
\label{P_A}
\end{align}
Clearly, this function is not normalizable since $P_A(F_z,\xi)\sim |F_z|^{-(1+\alpha)}$ for small force fields, hence it is not a PDF. It is easily verified with simple integration that Eq.~\eqref{P_A} gives the moments found in Eqs.~(\ref{moments},\ref{SeperateMoment}). Mathematically, we are interested in the limit where both $F_c\rightarrow\infty$ and $F_z\rightarrow\infty$, hence we denote $\tilde{F}=F_z/F_c$. This corresponds to the large forces in the problem. We present the solution using a natural scale, namely, we define
\begin{equation}
	\mathcal{I}_{\alpha}(\tilde{F})=F_c^{1+\alpha}P_A(F_z,\xi)
	\label{infinite}
\end{equation}
and $\mathcal{I}_{\alpha}(\tilde{F})$ is the infinite density of $\tilde{F}$, where the name comes from the fact that $\int_{-\infty}^{\infty} \mathcal{I}_{\alpha}(\tilde{F})d\tilde{F}=\infty$. With the usage of Eq.~\eqref{infinite} we find
\begin{eqnarray}
	\mathcal{I}_{\alpha}(\tilde{F})=
	\begin{cases}
		\frac{2\pi \alpha F_H^\alpha}{3(1+\alpha)}\left(\frac{1}{|\tilde{F}|^{1+\alpha}}-1\right) & |\tilde{F}|<1\\
		0 & |\tilde{F}|\geq 1.
	\end{cases}
	\label{44}
\end{eqnarray}
While we may use Eq.~\eqref{44} to obtain force moments, the remaining question is how can it be measured, at least in principle?
Clearly, $\mathcal{I}_\alpha(\tilde{F}) \sim \tilde{F}^{-(1+\alpha)}$ for small $\tilde{F}$, the question is what is the physical meaning of this non-normalized solution? Namely, how is the infinite density related to the normalized probability density of the force field.

We realize that moments of the force field can be obtained not only from the infinite density, instead we may use the probability density of the force itself, namely $P(F_z,\xi)$. From here, we reach the conclusion
\begin{equation}
	\mathcal{I}_\alpha(\tilde{F}) =\lim_{F_c,F_z\rightarrow\infty} F_c^{1+\alpha}P(F_z,\xi),
	\label{PDF}
\end{equation}
namely, the normalized density $P(F_z,\xi)$ is related to $\mathcal{I}_\alpha(\tilde{F})$. It is important to emphasize that while Eq.~\eqref{PDF} is an exact statement that holds in a limit, for finite $F_c$ the theory holds as a valid approximation, as we now demonstrate.

In numerous areas of physics, the observed tracer is typically small and fulfills the condition $\rho a^d\ll1$, indicating a significantly low density \cite{Von,Jean,Kador}. This condition holds true, for instance, in a two-dimensional system, such as a disk, where a tiny tracer with a small radius $a$ is positioned at its center. The tracer is encompassed by positively/negatively charged particles. Here we demonstrate this relation for the case studied by Holtsmark, where $a$ is finite though small.

We simulated the random force field for the case $\delta=2$ and $d=3$ and obtained $P(F_z,\xi)$, (see details in Appendix \ref{Appendix A}). Recall that $\delta =2$ implies a gravitational of Coloumb type or force fields. The results (dotted line) are compared with the theory in Fig.~\ref{fig:1}. After re-scaling, using Eq.~\eqref{PDF}, we compare numerical data to the exact one, namely, the inverse Fourier transform of Eq.~\eqref{d3Characteristic} (black solid line), and to $\mathcal{I}_ \alpha(\tilde{F})$ found in Eq.~\eqref{44} (dashed line). The figure demonstrates perfect agreement between statistics of the simulated field and the non normalized solution, $\mathcal{I}_\alpha(\tilde{F})$. Notice, for $\tilde{F}=1$, there is a cut-off indicating that the largest total force is of the order $F_c$, which is equivalent to the largest force exerted by a single particle in the vicinity of the tracer \cite{Vezzani,PHChavanis}. The next step is to consider another scaling solution of the problem, found when $a=0$, corresponding to the original work of Holtsmark. 

\subsection{L\'{e}vy-Holtsmark statistics}
The characteristic function, Eq.~\eqref{d3Characteristic}, for the case of $a=0$ is given by
\begin{eqnarray}
	\tilde{P}(k,0)=\exp\left[-\mu_{d,\alpha}|F_Hk|^\alpha\right],
	\label{8}
\end{eqnarray}
where $\mu_{d,\alpha}$ is a dimension-less constant given by
\begin{eqnarray}
	\label{9}
	\mu_{d,\alpha}=\pi\begin{cases} \frac{\Gamma\left(1-\frac{\alpha}{2}\right)}{2^\alpha\Gamma\left(1+\frac{\alpha}{2}\right)}& d=2\\ \frac{4\cos(\frac{\pi\alpha}{2})\Gamma(1-\alpha)}{3(1+\alpha)} & d=3.
	\end{cases}
\end{eqnarray}
Eq.~\eqref{8} is the Fourier transform of the well-known L\'{e}vy stable distribution function, $P(F_z,0)=L_\alpha(F_z)$, for $0<\alpha<2$. Here $L_\alpha (F_z)= L_\alpha(-F_z)$ from symmetry. From Eq.~\eqref{8} we see that  force scale defined above Eq.~\eqref{17}, $F_H$, determines the width of the distribution of the force field, when $a=0$, namely, $\xi=0$. For $d=3$ and $\delta=2$, namely $\alpha=3/2$, by applying the inverse Fourier transform over Eq.~\eqref{8}, we recover the Holtsmark distribution, which is a special case of the L\'{e}vy stable distribution. The following question arises: what is the connection between the L\'{e}vy-Holtsmark distribution and the infinite density? This question is answered next.

\subsection{Relation of infinite density and L\'{e}vy statistics}
Applying an inverse Fourier transform ($\mathcal {F}^{-1}$) on the characteristic function yields the L\'{e}vy distribution of $F_z$, where $P (F_z,0) =L_\alpha(F_z) =\mathcal{F}^{-1}\left [\exp\left (-\mu_{d, \alpha}|F_Hk|^\alpha\right)\right] $. The function $L_\alpha(F_z)$ is tabulated in programs like \textit{Mathematica} and hence easy to plot. We now notice that the L\'{e}vy density for large $F_z$ matches the solution we found here, namely, the infinite density for small $F_z$. We have by using the large $F_z$ limit of $L_\alpha(F_z)$ and the small $F_z$ limit of Eq.~\eqref{44} 
\begin{equation}
	\begin{aligned}
		L_\alpha(F_z)&\sim \frac{2\pi\alpha F_H^\alpha}{3(1+\alpha)}\frac{1}{|F_z|^{1+\alpha}}, \\P(F_z,\xi) &\sim \frac{\mathcal{I}_\alpha(F_z)}{F_c^{1+\alpha}}\sim \frac{2\pi\alpha F_H^\alpha}{3(1+\alpha)}\frac{1}{|F_z|^{1+\alpha}},
	\end{aligned}
\end{equation}
hence, the two solutions match as they should. In other words, the L\'{e}vy distribution accurately describes the center part of $P(F_z, \xi)$ in the limit of a small but finite $a$, whereas our solution accurately describes the large $F_z$ limit. As mentioned in the introduction, the study of large forces is crucial, and that regime is described by the infinite density found here.

\section{SHARP STATISTICAL TRANSITION}
We now show how the moments of the force field exhibit bi-linear scaling with a sharp transition found when the order of the moments is modified.

Consider the absolute value of the moment, denoted as $\langle |F_z|^q\rangle$, where $q$ gets any non-negative real value. Since the L\'{e}vy/Holtsmark method of calculating the moments fails for  $q > \alpha$, mathematically because the moments diverge in that regime and physically since the assumption of absence of excluded volume $a=0$ cannot be used, we employ the infinite density. Hence, $\langle |F_z|^q\rangle$ is given by the integral,
\begin{equation}
	\langle |F_z|^{q} \rangle \sim \frac{1}{F_c^{1+\alpha}} \int_{-F_c}^{F_c} |F_z|^{q} \mathcal{I}_{\alpha}(F_z) dF_z, \quad q>\alpha.
	\label{momentinf}
\end{equation}
Unlike Eq.~\eqref{SeperateMoment} now $q$ is not necessarily an integer. The moments for $q<\alpha$ cannot be calculated by the infinite density, since the latter does not describe well the small force fields and the normalization condition $q=0$ case, hence they are found using the L\'{e}vy's distribution. The infinite density and the L\'{e}vy distribution are complementary, thus each succeeds where the other one fails. Hence, $\langle |F_z|^q\rangle$ for $q<\alpha$ is obtained by solving the integral,
\begin{equation}
\langle |F_z|^q \rangle = \int_{-\infty}^{\infty} |F_z|^q L_{\alpha}(F_z) dF_z.
\end{equation}
The final solution for $\langle |F_z|^{q} \rangle$ is
\begin{equation}
	\langle |F_z|^q\rangle\sim
	\begin{cases}
		 M_{q>\alpha}F_H^{\alpha}F_c^{q-\alpha} & q>\alpha \\ 
		M_{q<\alpha}F_H^q & q<\alpha,
	\end{cases}
	\label{79}
\end{equation}
where the amplitudes $M_q$ are
\begin{equation}
	\begin{cases}
	M_{q>\alpha}=\frac{4\pi}{\delta(q-\alpha)(q+1)}\\
	M_{q<\alpha}=\frac{(\mu_{d,\alpha})^{\frac{q}{\alpha}}\Gamma\left(1-\frac{q}{\alpha}\right)}{\cos\left(\frac{\pi q}{2}\right)\Gamma(1-q)}.
	\end{cases}
	\label{Moment amplitude}
\end{equation}
\begin{figure}[H]
	\includegraphics[width=0.49\textwidth]{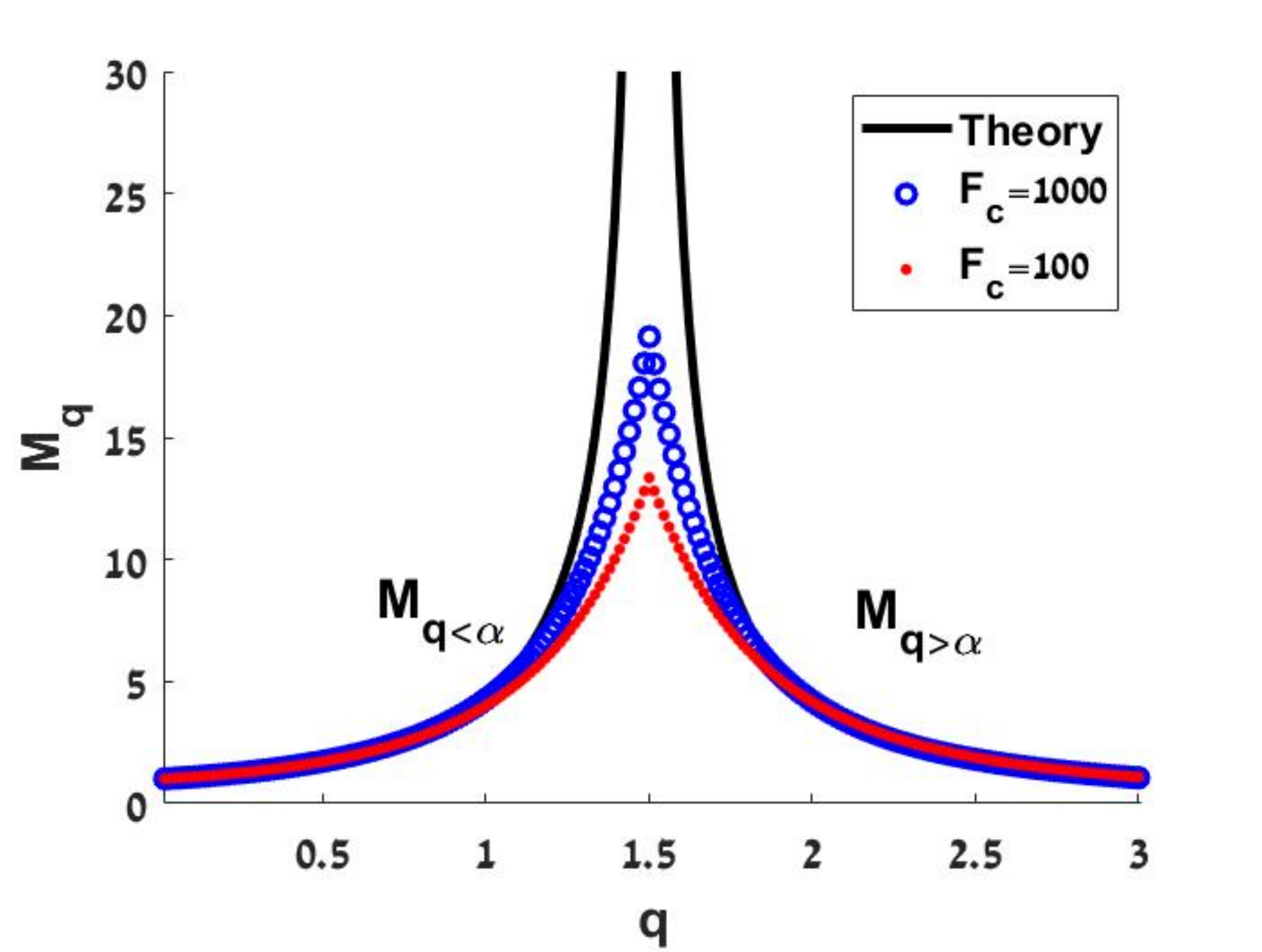}
	\includegraphics[width=0.49\textwidth]{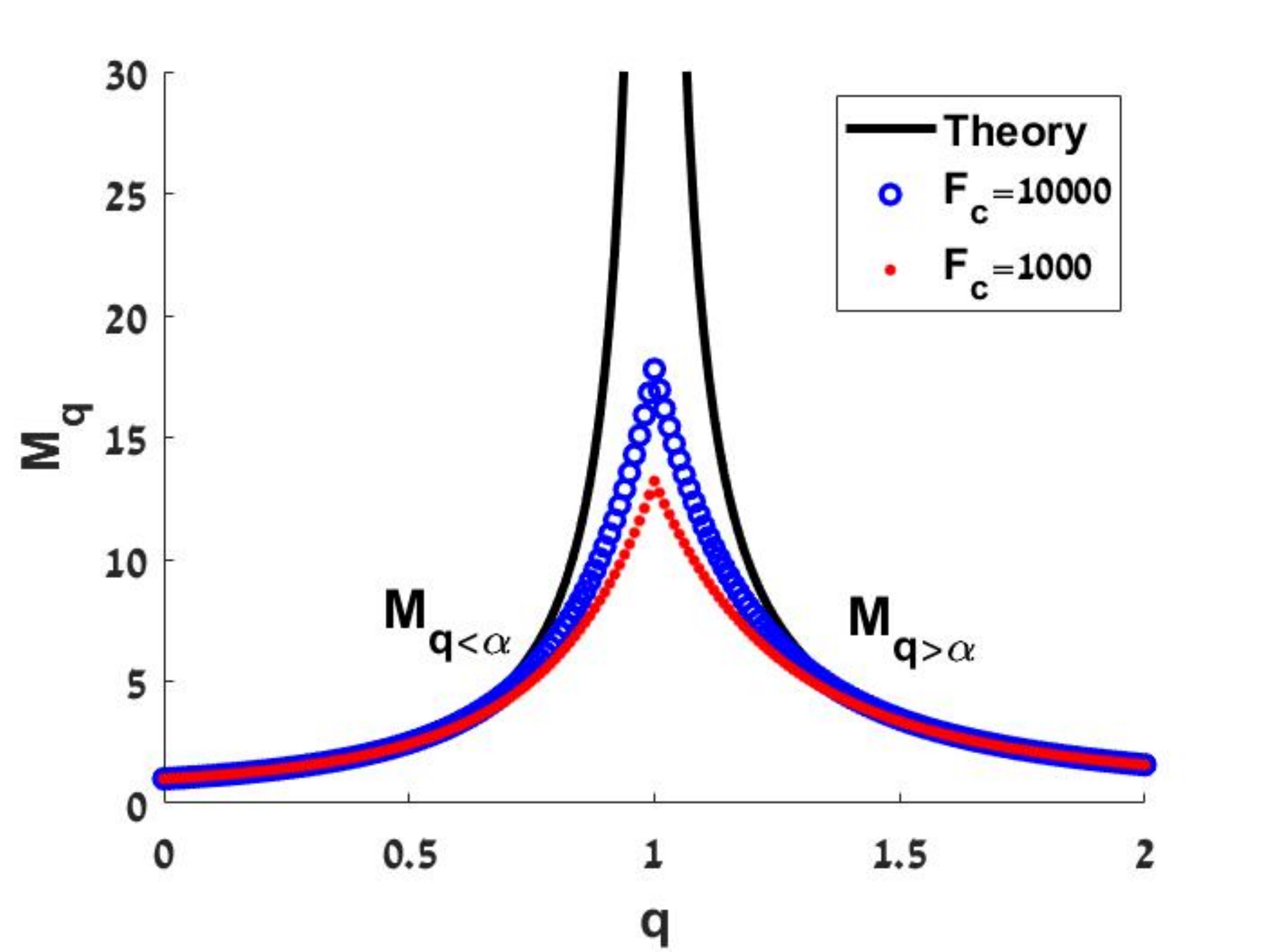}
	\caption{We present numerical results for the moments amplitude $M_q$ versus $q$ (shapes). In the limit of large $F_c$, these converge to the theoretical prediction Eq.~\eqref{Moment amplitude} and Eq.~\eqref{Momentamplitude} (see Appendix \ref{Appendix C}) for $d=3,2$, respectively. In the large limit of $F_c$, these moments diverges as $q\rightarrow\alpha$, as shown. At the top (bottom) graph $\rho=3/4\pi$ $(\rho=1/\pi)$, $\alpha=3/2$ $(\alpha=1)$, and $d=3$ $(d=2)$, respectively. In both graphs we chose $C=1$, such that $F_c=1/a^\delta$. The plots show how $q>\alpha$ corresponds to the infinite density scaling, while $q<\alpha$ to the Holtsmark-L\'{e}vy law.}
	\label{fig:3}
\end{figure}
\noindent
There is a clear divergence of $\langle |F_z|^q\rangle$ for $q\rightarrow\alpha$ from above and below, thus the moments exhibit a  transition, which is an indication of a transition between statistical laws of weak fields (L\'{e}vy/Holtsmark) and strong fields (infinite density).

\subsection{The infinite density for dimension two}

As in the three-dimensional case we start with the calculation of the moments $\langle F_z^{2n}\rangle$, where $n$ is a non-negative integer. With the usage of the characteristic function, found using Eq.~\eqref{4},
\begin{equation}
	\langle e^{ikF_z}\rangle=\exp\left[-\pi \xi^{\alpha}\left({}_1F_2\left[-\frac{\alpha}{2};1,1-\frac{\alpha}{2};-\frac{|F_ck|^2}{4}\right]-1\right)\right],
	\label{Char2D}
\end{equation}
and Eq.~\eqref{MGF}, the $2n$-th moment for the dimensionless variable $\tilde{F}=F_z/F_c$ is
\begin{equation}
	\langle\tilde{F}^{2n}\rangle\sim\frac{\sqrt{\pi}\alpha\xi^{\alpha}}{2n-\alpha}\left(\frac{2\Gamma(\alpha)}{\Gamma(\frac{\alpha}{2})}+\underbrace{\frac{\Gamma(n+\frac{1}{2})}{\Gamma(n+1)}-\frac{\Gamma\left(\frac{1+\alpha}{2}\right)}{\Gamma\left(1+\frac{\alpha}{2}\right)}}_{g_\alpha(n)}\right),
	\label{moment2D}
\end{equation}
where we used $\xi\ll1$ as before. In Appendix \ref{Appendix C}, we present the first three non-zero moments for $d=2$. Our next goal is to find the infinite density that generates the moments in Eq.~\eqref{moment2D}. For that aim, we use the Mellin transform, with Eq.~\eqref{Mellin} and the identity 
\begin{equation}
	 g_\alpha(n)=\frac{\alpha-2n}{\sqrt{\pi}(1+\alpha)}\int_{-1}^{1}{}_2F_1\left[\frac{1}{2},\frac{1+\alpha}{2};\frac{3+\alpha}{2},\tilde{F}^2\right]\tilde{F}^{2n}d\tilde{F},
\end{equation}
where $g_\alpha(n)$ is defined in Eq.~\eqref{moment2D}. Therefore, the infinite density is
\begin{align}
	&\mathcal{I}_{\alpha}(\tilde{F})=\nonumber\\&\begin{cases}
	\alpha F_H^\alpha \left[\frac{\sqrt{\pi}\Gamma\left(\frac{1+\alpha}{2}\right)}{\alpha\Gamma\left(\frac{\alpha}{2}\right)|\tilde{F}|^{1+\alpha}}-\frac{{}_2F_1\left[\frac{1}{2},\frac{1+\alpha}{2};\frac{3+\alpha}{2},\tilde{F}^2\right]}{(1+\alpha)}\right] & |\tilde{F}| < 1 \\
	0 & |\tilde{F}| \geq 1,
\end{cases}
	\label{infdens2D}
\end{align}
where ${}_2F_1$ is the Gaussian hypergeometric function \cite{Hyper}. Of course, one may insert Eq.~\eqref{infdens2D} in Eq.~\eqref{momentinf}, with $q=2n$ and verify Eq.~\eqref{moment2D}. The infinite density in two and three dimension are clearly different from each other, but both satisfy the relations in Eqs.~(\ref{infinite}, \ref{PDF}). Still in both dimensions the infinite density has similar behavior for $\tilde{F}\rightarrow0$, \textit{i.e.}, $\mathcal{I_\alpha}(\tilde{F})\sim\tilde{F}^{-(1+\alpha)}$, so this is a non-normalized function.

In Fig~\ref{fig:3}, we compare the theoretical result of the amplitudes of the absolute value of the force moments, obtained from Eq.~\eqref{Moment amplitude} and Eq.~\eqref{infdens2D} (see Appendix \ref{Appendix C}), with the simulation for the Holtsmark ($d=3$, $\alpha=3/2$) and Cauchy ($d=2$, $\alpha=1$) cases, respectively. From this figure we see that for $q>\alpha$, the moments are obtained by the infinite density function, and for $q<\alpha$ by the L\'{e}vy distribution. As $F_c$ gets larger ($a$ smaller) so does the peak at $q=\alpha$, hence the data converges to the theoretical result. The figures clearly illustrate that by studying different orders of moments, we reveal different scales of the problem, accompanied by a sharp transition, found at $q_c=\alpha$.

\subsection{Extension of this work}
In many stochastic models, the noise is described by L\'{e}vy statistics \cite{Ariga}. L\'{e}vy noise cannot be realized in physical systems in its exact mathematical form, instead semi-truncated L\'{e}vy noise is used, for various far from equilibrium systems, including active swimmer suspension \cite{Dais,Micro,Kuri}, actomyosin networks \cite{Network}, and cultured cell \cite{Cultured}. Here we showed using a static model, that indeed the forces are truncated, and that this truncation is related to finite size effects, namely to the radius $a$, and more importantly this cutoff is at least in a static description deeply related to the infinite density concept. Since the far tail of the distribution of the random force is important for the enhancement of active diffusion, our work may impact the whole field. The remaining challenge is to see how the statistical laws found here for a basic static model translate into a dynamical picture \cite{Kana}.

\section{SUMMARY}
To conclude, we found a non-trivial behavior of the moments of the force field for Holtsmark-like problems. A transition controlled by the order of the moments is observed at a critical value of $q_c=\alpha=d/\delta$. The $q$ moments of order $q>q_c$ are described by the cutoff force scale $F_c$, which is determined by a single bath particle in the vicinity of the tracer, so $\langle|F|^q \rangle \propto (F_c)^{q-\alpha}$. In contrast, low order moments $q<q_c$ are given by the Holtsmark force scale. The amplitudes of these moments, $M_q$, diverge in the vicinity of the transition point $q_c$. Further, the low order moments are determined by the L\'evy-Holtsmark law, while the higher order moments, namely $q>q_c$, are determined by the infinite density found here. The L\'{e}vy-Holtsmark distribution and the infinite density are complementary scaling laws of the problem. The infinite density in Eqs.~(\ref{infdens2D},\ref{44}) describes the distribution of forces, $F_z$, for large forces. These are in many applications important, as large forces can lead to violent effects. More mathematically, in the limit where both $F_c$ and $F_z$ are large, we get a limit theorem that is complimentary to the well-known Holtsmark distribution. This is found using the small density limit, where the assumptions of the model are valid. The PDF of forces, when properly re-scaled, yields the infinite density, and importantly, the latter describes the large forces in the problem (see Fig.~\ref{fig:1}). As such, the infinite density is an essential part of this problem, exactly like the well known L\'{e}vy-Holtsmark distribution.

\section*{ACKNOWLEDGMENTS}
	This work was supported by the Israel Science Foundation grant 1614/21 

\begin{appendices}
\section{SIMULATION OF THE MODEL \label{Appendix A}}
Here we give an explanation about the simulation of the model. We scattered uniformly $N$ particles in a $d=2, 3$ dimensional sphere with outer radius denoted as $R$. The force each particle exerts on a rigid body tracer with a radius of $a$ located in the center is then measured (bath particles are excluded from the volume with radius $a\ll R$). By summing all the observed forces, one obtains the total force field applied on the rigid tracer presented in Eq.~\eqref{3}. By repeating this process many times, the distribution of the total force, $P(F_z,\rho a^d)$, is obtained. We also obtained the PDF by performing numerical inverse Fourier transform of Eq.~\eqref{4}, using \textit{Mathematica}. In both Fig.~\ref{fig:1} and Fig.~\ref{fig:3}, $R=20$, which is much greater than $a$.
\section{DERIVATION OF THE FORCE MOMENTS IN THREE DIMENSION \label{Appendix B}}
The exact result of the force moments in dimension $d=3$ are given here. Since the odd moments are vanishing, we deal only with the even ones, starting with the variance. We expand the Hyper geometric function presented in Eq.~\eqref{d3Characteristic} as a Taylor series and find
\begin{equation}
	\langle e^{ikF_z}\rangle=\exp(\sum_{n=1}^{\infty}\frac{4\pi\alpha \xi^\alpha F_c^{2n}}{3(2n-\alpha)(2n+1)}\frac{(ik)^{2n}}{(2n)!}).
	\label{S1}
\end{equation}
The expansion of Eq.~\eqref{S1} in a Taylor series,
\begin{equation}
	\langle e^{ikF_z}\rangle=1+\sum_{m=1}^{\infty}\frac{1}{m!}\left(\sum_{n=1}^{\infty}\frac{4\pi\alpha 	\xi^\alpha F_c^{2n}}{3(2n-\alpha)(2n+1)}\frac{(ik)^{2n}}{(2n)!}\right)^m,
	\label{S2}
\end{equation}
with the usage of Eq.~\eqref{MGF}, gives all the exact moments. We provided only the first three non-zero exact moments in the paper.
\section{DERIVATION OF THE FORCE MOMENTS IN TWO DIMENSION \label{Appendix C}}
We start by expanding Eq.~\eqref{Char2D} in a Taylor series
\begin{equation}
\langle e^{ikF_z}\rangle = 1+\sum_{m=1}^{\infty}\frac{1}{m!}\left(\sum_{n=1}^{\infty}\frac{\alpha\sqrt{\pi}\xi^\alpha F_c^{2n}\Gamma(n+\frac{1}{2})}{(2n-\alpha)\Gamma(n+1)}\frac{(ik)^{2n}}{(2n)!}\right)^m.
\label{Expands2D}
\end{equation}
Using Eqs.~(\ref{MGF},\ref{Expands2D}) , the variance is
\begin{equation}
	\langle F_z^2\rangle =\frac{\pi \alpha \xi^\alpha}{2(2-\alpha)}F_c^2.
\end{equation}
We find the fourth moment, $\langle F_z^4\rangle$, and the sixth moments, $\langle F_z^6\rangle$, in the same way
\begin{equation}
	\langle F_z^{4}\rangle =\left[\frac{3\pi\alpha\xi^{\alpha} }{8(4-\alpha)}+3\left(\frac{\alpha\pi\xi^\alpha}{2(2-\alpha)}\right)^2\right]F_c^4,
	\label{F4}
\end{equation}
\begin{equation}
	\begin{aligned}
	&\langle F_z^{6}\rangle =&\\ &\left[\frac{5\pi\alpha\xi^{\alpha} }{16(6-\alpha)}+\frac{45\alpha^2\pi^2\xi^{2\alpha}}{16(4-\alpha)(2-\alpha)}+15\left(\frac{\alpha\pi\xi^\alpha}{2(2-\alpha)}\right)^3\right]F_c^6.
	\end{aligned}
	\label{F6}
\end{equation}
For $\xi\ll1$, the leading term in Eq.~\eqref{F4} is
\begin{equation}
	\langle F_z^{4}\rangle \sim \frac{3\pi\alpha\xi^\alpha}{8(4-\alpha)}F_c^4.
	\label{F4asymptotic}
\end{equation}
Similarly, the $2n$-th moment is
\begin{equation}
	\langle F_z^{2n}\rangle =\frac{\alpha\sqrt{\pi}\xi^{\alpha}\Gamma\left(n+\frac{1}{2}\right) }{(2n-\alpha)\Gamma(n+1)}F_c^{2n}+O(\xi^{2\alpha}).
	\label{2Dmoments}
\end{equation}
Using the rescaled variable $\tilde{F}=F_z/F_c$, we get
\begin{equation}
	\langle \tilde{F}^{2n}\rangle =\frac{\alpha\sqrt{\pi}\xi^{\alpha}\Gamma\left(n+\frac{1}{2}\right) }{(2n-\alpha)\Gamma(n+1)}+O(\xi^{2\alpha}),
\end{equation}
which yields Eq.~\eqref{moment2D}.

Our next goal is to find the moments of the absolute force value that are mentioned in Eq.~\eqref{79}, $\langle |F_z|^q\rangle$. We do so by employing the infinite density presented in Eq.~\eqref{infdens2D} for $q > \alpha$ and $L_\alpha(F_z)$ for the moments $q<\alpha$, using $\langle |F_z|^q \rangle = \int_{-\infty}^{\infty} |F_z|^q L_{\alpha}(F_z) dF_z$, which yield Eq.~\eqref{79}. The amplitudes, $M_q$, for the case of $d=2$ are
\begin{equation}
	\begin{cases}
		M_{q>\alpha}=\frac{\alpha\sqrt{\pi}\xi^{\alpha}\Gamma\left(\frac{q+1}{2}\right) }{(q-\alpha)\Gamma\left(1+\frac{q}{2}\right)},\\
		M_{q<\alpha}=\frac{(\mu_{d,\alpha})^{\frac{q}{\alpha}}\Gamma\left(1-\frac{q}{\alpha}\right)}{\cos\left(\frac{\pi q}{2}\right)\Gamma(1-q)},
	\end{cases}
	\label{Momentamplitude}
\end{equation}
with $\mu_{d,\alpha}$ defined in Eq.~\eqref{9}. Eq.~\eqref{Momentamplitude} is used to describe how the finite size simulations converge to the asymptotic prediction in Fig.~\ref{fig:3} presented in the letter.\\[5cm]
\end{appendices}


\begin{thebibliography}{99}
\bibitem{Jean}
J. Holtsmark, {\em Ann. Physik}, {\bf 58}, 577 (1919).
\bibitem{Von}
S. Chandrasekhar, and J. Von Neumann, {\em Astrophysical Jouranl}, {\bf 95} (1942).
\bibitem{Chandrash}
S. Chandrasekhar, \textit{Stochastic Problems in Physics and Astronomy}, {\em Rev, Mod. Phys}, {\bf 15}, 1-89 (1943).
\bibitem{Figueiredo}
A. D. Figueiredo, T. M. da Rocha Filho, and M. A. Amato, {\em J. Math. Phys.} {\bf 60}, 073301 (2019).
\bibitem{Piet}
L. Pietronero, M. Bottaccio, R. Mohayaee, and M. Montuori, {\em J. Phys.} {\bf 14}, 9 (2002).
\bibitem{Wes}
J. H. Wesenberg, and K. Molmer, {\em Phys. Rev. Lett.} {\bf 93}, 143903 (2004).
\bibitem{Gel}
D. W. Swarts, and B. A. Camley, {\em Soft Matter} {\bf 17}, 9876-9892 (2021).
\bibitem{Dunkel}
I. M. Zaid, J. Dunkel and J. M. Yeomans, {\em J. Royal Soc. Interface} {\bf 8}, 1314–1331 (2011).
\bibitem{Klafter}
J. Klafter, and I.M. Sokolov, \textit{First Steps in Random Walks: From Tools to Applications}, {\em Oxford University Press} (2011).
\bibitem{Kolmogorov}
B. V. Gnedenko and A. N. Kolmogorov, \textit{Limit Distribution for Sum of Independent Random Variables}, {\em Cambridge University Press} (1954).
\bibitem{Hazut}
N. Hazut, S. Medalion, D.A. Kessler, and E. Barkai, {\em Phys. Rev. E.} {\bf 91}, 052124  (2015).
\bibitem{Zarfaty}
L. Zarfaty, A. Peletskyi, E. Barkai, and S. Denisov, {\em Phys. Rev. E}
{\bf 100}, 042140 (2019).
\bibitem{Sire}
P.H. Chavanis, and C. Sire, {\em Phys. Rev. E} {\bf 62}, 490-506 (2000).
\bibitem{Bruce}
B. J. West, and W. Deering, {\em Phys. Report} {\bf 246}, issue 1-2 (1994).
\bibitem{Biel}
A. Bielinskyi, S. Semerikov, V. Solovieva, and V. Soloviev, {\em EDP Sciences} {\bf 65}, 06006 (2019).
\bibitem{Kuri}
T. Kurihara, M. Aridome, H. Ayade, I. Zaid and D. Mizuno, {\em Phys. Rev. E} {\bf 95}, 030601(R) (2017).
\bibitem{Plasma}
G. Zimbardo and S. Perri, {\em Astrophys. J.} {\bf 778}, 35 (2013).
\bibitem{finiteuniverse}
S. Heath and L. Shepp, {\em A Garden of Quanta by World Sci. Publ., River Edge, NJ}, 155-166 (2003).
\bibitem{Dais}
I. Zaid, and D. Mizuno, {\em Phys. Rev. Lett} {\bf 117}, 030602 (2016).
\bibitem{Kador}
E. Barkai, A. V. Naumov, Yu. G. Vainer, M. Bauer, and L. Kador, {\em Phys. Rev. Let.} {\bf 91}, 075502 (2003).
\bibitem{Barkai}
E. Barkai, R. Silbey, and G. Zumofen, {\em Phys. Rev. Lett} {\bf 84}, 5339 (2000).
\bibitem{Stoneham}
A. M. Stoneham, {\em Rev. Mod. Phys.} {\bf 41}, 82 (1969).
\bibitem{Davidson}
G. Afek, N. Davidson, D. A. Kessler, and E. Barkai, arXiv:2107.09526, {\em cond-mat.stat-mech}(2021) {\em Review of Modern Physics}(in press).
\bibitem{Wan}
P. Xu, R. Metzler, and W. Wang, {\em Phys. Rev. E} {\bf 105}, 044118 (14 April 2022).
\bibitem{infdens}
V. Zaburdaev, S. Denisov, and J. Klafter, {\em  Rev. Mod. Phys.} {\bf 87}, 483 (2015).
\bibitem{Kesler}
E. Barkai, G. Radons, and T. Akimoto, {\em Phys. Rev. Lett.}, {\bf 127}, 140605 (2021).
\bibitem{Erez}
E. Aghion, D. A. Kessler, E. Barkai, {\em Phys. Rev. Lett.} {\bf 122}, 010601 (2019).
\bibitem{EliB}
A. Rebenshtok, S. Denisov, P. H\"{a}nggi, and E. Barkai, {\em Phys. Rev. E} {\bf 90}, 062135 (2014).
\bibitem{Leibovich}
N. Leibovich, and E. Barkai, {\em Phys. Rev. E}  
{\bf 99}, 042138 (2019). 
\bibitem{Ralf}
S. Giordano, F. Cleri, and R. Blossey, {\em Phys. Rev. E} {\bf 107}, 044111 (2023)
\bibitem{Radons}
T. Akimoto,  E. Barkai, and  G. Radons, {\em Phys. Rev. E.} {\bf 105}, 064126 (2022).
\bibitem{Aaronson}
J. Aaronson, {\em Mathematical Surveys and Monographs} {\bf 50} (1997).
\bibitem{Invariant}
N. Korabel, and E. Barkai, {\em Phys. Rev. Lett.} {\bf 102}, 050601 (2009).
\bibitem{Kana}
K. Kanazawa, T. G. Sano, A. Cairoli, and A. Baule, {\em Nature} {\bf 579}, 7799 (2020).
\bibitem{Micro}
D. L. Koch, and G. Subramanian, {\em Rev. Fluid Mech.} {\bf 43}, 637–659 (2011).
\bibitem{Jansen}
G. H. Jansen, \textit{Coulomb interactions in particle beams}, {\em  Nucl. Instrum. Methods Phys. Res., Sect. A}, {\bf 298}, 496 (1990).
\bibitem{Strong-anomalous}
P. Castiglione, A. Mazzino, P. Muratore-Ginanneschi, and A. Vulpiani, {\em Physica D: Nonlinear Phenomena}, {\bf 134}, 1 (1999).
\bibitem{anomalous in living cell}
N. Gal, and D. Weihs, {\em Phys. Rev. E}, {\bf 81}, 020903(R) (2010).
\bibitem{Displacement}
J. Vollmer, L. Rondoni, M. Tayyab, C. Giberti, and C. Mej\'{i}a-Monasterio, {\em Phy. Rev. Research}, {\bf 3}, 013067 (2021).
\bibitem{Martin}
T. Dauxois, S. Ruffo, E. Arimondo, and M. Wilkens, {\em Lect. Notes Phys.} {\bf 602}, 1-19 (23 Aug 2002).
\bibitem{RebePRL}
A. Rebenshtok, S. Denisov, P. H\"anggi, and E. Barkai, {\em Phys. Rev. Letters} {\bf 112}, 110601 (2014).

\bibitem{Hyper}
M. Abramowitz, and I. A. Stegun, \textit{Handbook of Mathematical Functions With Formulas, Graphs, and Mathematical Tables}, {\em Dover, Washington, D.C.} (1972).
\bibitem{MellinTrans}
J. Bertrand, P. Bertrand, J. P. Ovarlez, \textit{The Mellin Transform. The Transforms and Applications Handbook: Second Edition}
{\em Boca Raton: CRC Press LLC} (2000)
\bibitem{Vezzani}
A. Vezzani, E. Barkai, and R. Burioni, {\em Phys. Rev. E} {\bf 100}, 012108 (2019).
\bibitem{PHChavanis}
P.H. Chavanis, {\em The European Physical Journal B}, {\bf 70}, 413-433 (2009).
\bibitem{Ariga}
T. Ariga, K. Tateishi, M. Tomishige, and D. Mizuno, {\em Phys. Rev. Lett.} {\bf 127}, 178101 (2021).
\bibitem{Network}
I. Zaid, H. L. Ayade, and D. Mizuno, {\em Biophys. J.} {\bf 106}, 171a (2014).
\bibitem{Cultured}
É. Fodor, M. Guo, N. S. Gov, P. Visco, D. A. Weitz, and F.
van Wijland, {\em Europhys. Lett.} {\bf 110}, 48005 (2015).





	\end{thebibliography}
\end{document}